\documentclass[a4paper,11pt]{article}
\pdfoutput=1 

\usepackage{jinstpub} 

\usepackage{lineno}
\usepackage{amssymb}
\usepackage{amsmath}
\usepackage{url}
\usepackage[utf8]{inputenc}
\usepackage{ifthen}
\usepackage{ltxtable}
\usepackage{subfigure}
\usepackage{xspace}
\usepackage{here}
\usepackage{hyperref}
\usepackage{ulem}  

\title{\boldmath Classification and Recovery of Radio Signals from Cosmic Ray Induced Air Showers with Deep Learning}

\author{M. Erdmann,}
\author{F. Schl\"uter\footnote{Now at Karlsruhe Institute of Technology, Germany}}
\author{and R. \v{S}m\'{i}da\footnote{Now at Kavli Institute for Cosmological Physics and the Enrico Fermi Institute, The University of Chicago, 5640 S. Ellis Ave, Chicago, Il 60637, USA}}
\affiliation{Physics Institute 3A, RWTH Aachen University, Otto-Blumenthal-Str., 52056 Aachen, Germany}

\emailAdd{smida@kicp.uchicago.edu}

\abstract{Radio emission from air showers enables measurements of cosmic particle kinematics and identity. The radio signals are detected in broadband Megahertz antennas among continuous background noise. We present two deep learning concepts and their performance when applied to simulated data. The first network classifies time traces as signal or background. We achieve a true positive rate of about $90\%$ for signal-to-noise ratios larger than three with a false positive rate below $0.2\%$. The other network is used to clean the time trace from background and to recover the radio time trace originating from an air shower. Here we achieve a resolution in the energy contained in the trace of about $20\%$ without a bias for $80\%$ of the traces with a signal. The obtained frequency spectrum is cleaned from signals of radio frequency interference and shows the expected shape.}


\def\Offline{\mbox{$\overline{\rm Off}$\hspace{.05em}\raisebox{.3ex}{$\underline{\rm line}$}}\xspace}

\begin{document}
\maketitle
\flushbottom

\section{Introduction}

In modern experimental setups, sensors continuously convert physical signals to electric charge. 
Continuous data analysis is required to determine whether data should be saved or discarded.
Ideally, an analysis will recover a physical signal immediately after being recorded by the sensor, thereby minimizing the bandwidth required for data transfer. To distinguish the desired signal information from background noise, this live data analysis needs to be fast, have high signal selection efficiency, and a low rate of false positive decisions on background.

In this work, we investigate strategies for solving such challenges with deep learning techniques. These methods are based on neural networks with a substantial number of adjustable parameters to accommodate, for example, analyses of signal shapes. 
We take advantage of several network architectures and methods developed in the field of computer science which have been shown to be capable of handling millions of parameters \cite{Hinton, autoencoder, Ciresan, ILSVRC2012, ResNet1}. For a review on deep learning techniques refer to~\cite{review}. 
Recent applications in various particle and astroparticle research projects have demonstrated advantages when using deep learning methods \cite{NoVA, exotics, higgs, kaggle, jetflavor, Baldi:2016fql, 2017PhRvD..96g4034S, ttH,deOliveira:2017pjk, Erdmann:2017str, Paganini:2017hrr, Paganini:2017dwg, Erdmann:2018kuh, Erdmann:2018jxd}.

Our example applications focus on radio signals emitted by ultra-high energy cosmic rays (UHECRs) which initiate extensive particle showers in the atmosphere. UHECRs are likely protons and nuclei with energies extending from $10^{18}$~eV to above $10^{20}$~eV. The UHECR flux decreases from $63$~(km$^2$~yr~sr)$^{-1}$ at $10^{18}$~eV to only $10^{-3}$~(km$^2$~yr~sr)$^{-1}$ at $10^{20}$~eV~\cite{auger-spectrum1, Fenu2017}. Due to their electric charge and propagation through interstellar and intergalactic magnetic fields, no point source has been identified so far. To overcome these challenges, one looks for precise, reliable and low-cost measurement techniques, capable of achieving exposures well above ten thousand km$^2$~sr~yr. By accumulating sufficient statistics and having a precise reconstruction of a UHECR’s arrival direction, energy and mass composition, one can search for an excess of arrival directions around selected astronomical objects in the sky, with an angular offset related to the particle’s electric charge. Another approach is to search for ultra-high energy neutral particles, like photons and neutrinos, and use them for the identification of point sources. In both cases, one would need a solid triggering system followed by precise reconstruction of measured air showers. 

UHECRs are measured through the cascade of secondary particles they produce upon interaction with the earth’s atmosphere. This “extensive air shower” can cover a ground area of a few tens of km$^2$, and its electromagnetic component causes the emission of both fluorescence light along its trajectory, along with coherent radio signals. The coherent radio emission of shower particles leads to a transient radio signal with a duration of approximately $10^{-8}$~s. Such signals have been measured using broadband antennas, typically covering the frequency range $30-80$~MHz \cite{Geo-KahnLerche, Geo-Allan, LOPES, Codalema, AERA,TRex, LOFAR}, for corresponding cosmic-ray energies above $10^{17}$~eV~\cite{LOPES-Energy, LOFAR-Energy, TREX-Energy, AERA-Energy-PRL, AERA-Energy-PRD}. It has been proven that air shower signals can also be measured in the GHz frequency range~\cite{CROME}. For reviews on radio detection techniques refer to~\cite{Tim, Frank}.

The radio signal amplitude has been observed to increase quadratically with cosmic ray energy, and to decrease with the distance from the shower axis~\cite{LOPES-Energy, AERA-Energy-PRD, Radio}. The shape of the frequency spectrum of a radio signal exhibits a dependence on the height of the maximum of the particle shower development in the atmosphere~\cite{gate}. Therefore, both the signal magnitude and the frequency spectrum contain important information that needs to be recovered.

An irreducible background is present due to both Galactic radio signals and a range of terrestrial sources of radio emission. The background rate exceeds the cosmic ray rate by several orders of magnitude, which is of order $1$~Hz/\mbox{km$^2$}\ above $\sim10^{16}$~eV. In principle, the background rate can be reduced by a discriminator threshold used in a measuring sensor's output, but a trade-off would then then arise between background reduction and detection of low-amplitude radio signals from either low energy or distant cosmic ray showers. This challenging regime will be the primary interest in our investigations below.

Our objectives in this paper are twofold. First, we present a method to select cosmic ray induced radio signals while rejecting a large fraction of the radio background. For this classification task, a convolutional neural network is used. Second, we study a method to disentangle the cosmic ray radio signal from the simultaneously recorded noise. The aim is to conserve the radio signal with its original magnitude and frequency spectrum as accurately as possible. For this regression task, we use a bottleneck-like architecture similar to the de-noising autoencoder \cite{review} and we train it in a supervised manner. Similar independent work is described in \cite{Fuhrer:2018euq, arXiv:1812.03347}.

The data sets used for training and testing network models are described in section~\ref{Data}. Details of our classification and regression architectures are then provided in sections~\ref{Classification} and~\ref{Signal-recovery}, respectively. Finally, we present our conclusions in section~\ref{sec:conclusion}.

\section{Data sets of radio signals and noise}
\label{Data}

To train and test deep learning models we use large data sets of simulated radio signals and noise. Here we provide further details on the dataset and software.

\subsection{High-quality dataset of radio signals}

In order to have a high-quality dataset of radio signals we have adopted simulations made for the Auger Engineering Radio Array (AERA). AERA is a system of radio antennas installed at the Pierre Auger Observatory~\cite{Auger}, measuring pulses of a few nanoseconds in length emitted by cosmic ray air showers with energies above $10^{17}$~eV. More than $150$ autonomous antenna stations are spread over $17$~km$^2$ at various distances from each other. Each antenna consists of two horizontally polarized antenna systems, and is sensitive to frequencies from $30$ to $80$~MHz, with signal processing algorithms and electronics specifically developed for this purpose. Radio pulses are sampled at $180$~MHz and stored in traces of 1000 time bins in length. For this study we have chosen to analyze signals from AERA's logarithmic periodic dipole antennas (LPDAs).

Air showers were simulated using the CoREAS code~\cite{CoREAS}. CoREAS is a CORSIKA-based~\cite{CORSIKA} program, suitable for simulating the radio emission from air showers. It relies on the shower's particle content as simulated by CORSIKA, and calculates emitted electromagnetic radiation for each charged particle via the endpoint formalism~\cite{endpoint}. The showers were generated with the following parameters: zenith angle $\theta \in [0^\circ, 62^\circ]$ randomly sampled from a $\sin(\theta)\cos(\theta)$ distribution, azimuth angle $\phi$ uniformly distributed from $0^{\circ}$ to $360^{\circ}$, energy sampled uniformly in $\log_{10}(E)$ from $10^{17}$ to $10^{19}$~eV, and a randomly chosen shower core location.

The simulated air showers were further processed by the Pierre Auger Observatory's reconstruction pipeline, \Offline~\cite{Offline}. The signal was folded with the antenna response and converted to a voltage time trace. We have analysed the two station polarizations, East-West and North-South, independently in this work.

\subsection{Signal-to-noise ratio}
\label{Signal-to-noise ratio}

We characterize the strength of a signal in a "noisy" trace using the signal-to-noise ratio (SNR). The SNR is defined as the maximum amplitude of a signal divided by the root mean square (RMS) of the noise in the time trace
\begin{equation}
  \label{eq:snr}
  \text{SNR} = \frac{\text{max. Signal}}{\text{RMS}_{\text{Noise}}}
             = \frac{A_{\text{max}}^S}{\sqrt{\frac{1}{N}\sum^N_{i} A_i^2}}  \;,
\end{equation}
where $A_i$ represents the voltage in the $i$-th time bin of a trace of length $N$ bins, and $A_{\text{max}}^S$ is the maximum voltage of the signal. In the signal recovery task, the trace is divided into a signal and noise region and $A_{\text{max}}^S$ is calculated for the signal region, while the RMS noise is obtained from the noise region only. In the classification case, $A_{\text{max}}^S$ is determined from the true signal trace.

Let us note that the Pierre Auger Collaboration uses a different definition of the SNR for the AERA data, in which Eq.~\ref{eq:snr} is squared and convolved with a Hilbert envelope to calculate the maximum amplitude. This procedure increases the maximum amplitude of a signal by $\sim7\%$. The AERA data are typically reconstructed above $\text{SNR}_{\text{AERA}}=10$, which corresponds to $\text{SNR}\simeq\sqrt{10}\simeq3.2$ in our case.

In addition, air showers measured with AERA are externally triggered by the surface detector of the Pierre Auger Observatory~\cite{Auger}. The SNR is determined offline for each station-level signal. The signal is the sum of the electric field traces from all three polarizations: East-West, North-South, and vertical. The electric field trace is determined by unfolding the known antenna response. In our case, we examine the signal of each polarization independently, and the measured signal considered in the voltage trace includes antenna effects such as dispersion. Therefore, the signal has a lower SNR in our study when compared to the standard AERA analysis.

\subsection{Simulated data and noise}

We have prepared a code capable of generating various noise traces. This code provides maximal flexibility, as any input parameter can be modified and any radio component can be included. In this way we can debug and train our model on the Monte Carlo generated noise data by mimicking realistic radio background conditions at any site and using any electronic readout system.

Noise is simulated in frequency space and consists of two uncorrelated components of the same amplitude. The first component is white noise sampled from the normal distribution, i.e. voltage $V(f)=\mathrm{const.}$, where $f$ is frequency. The second component is colored noise, $V(f)\propto f^{-\alpha/2}$, where the power density index $\alpha$ takes a random value between $0$ and $1$ drawn from a uniform distribution. These two components mimic various sources of noise, and their sum agrees well with measurements (see e.g.~\cite{Fliescher}). In addition, any radio frequency interference (RFI) source emitting at a given frequency, or a transient source, can be included.

The last step is passing the sum of both noise components through a band-pass filter. We use the frequency band between $30$ and $80$~MHz to mimic the configuration of an AERA LPDA antenna. For the band-pass filter we use a finite impulse response (FIR) filter implemented in the SciPy package.

\paragraph*{Data for classification\\}

We normalize each generated time trace by dividing each bin by the standard deviation of the whole trace, before using it in a deep neural network. Normalization is known to improve the learning efficiency of a neural network. To study the performance of deep neural networks for different values of the SNR, we scale simulated signals relative to the standard deviation of the noise amplitude. For the classification study, the SNR ranges from $0.5$ to $5.0$ to cover an interesting range of signals for triggering. Low SNR values will show what fraction of weak signals can be identified while avoiding an excessive number of false positives. The upper bound of SNR values will be used to verify that a neural network correctly identifies large air shower signals. $\text{SNR}\,>\,3$ is typically used for the reconstruction of air showers (see \ref{Signal-to-noise ratio}).

\begin{figure}[ht]
	\centering
    \resizebox{0.8\columnwidth}{!}{\includegraphics{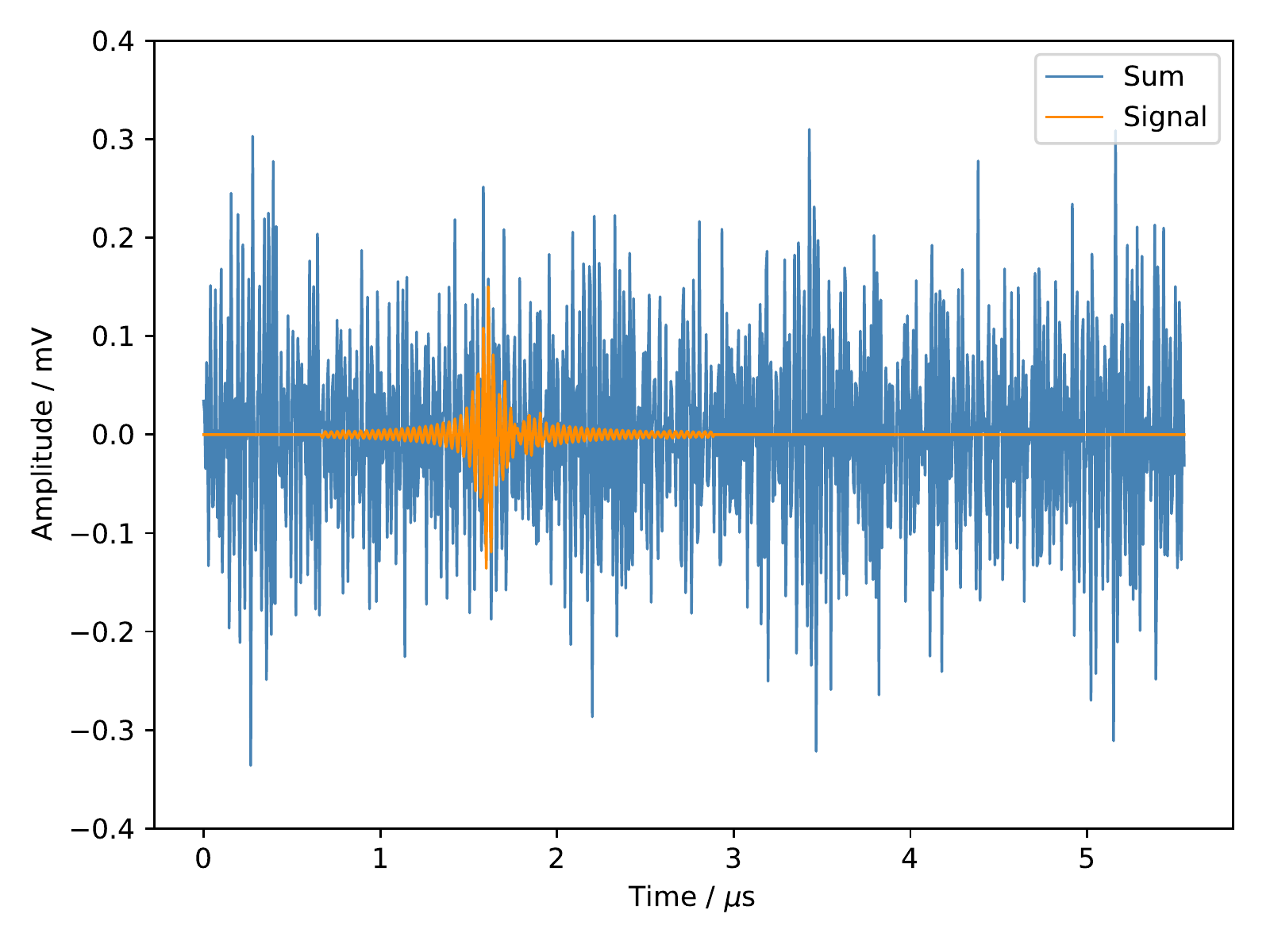}}\\
    \resizebox{0.8\columnwidth}{!}{\includegraphics{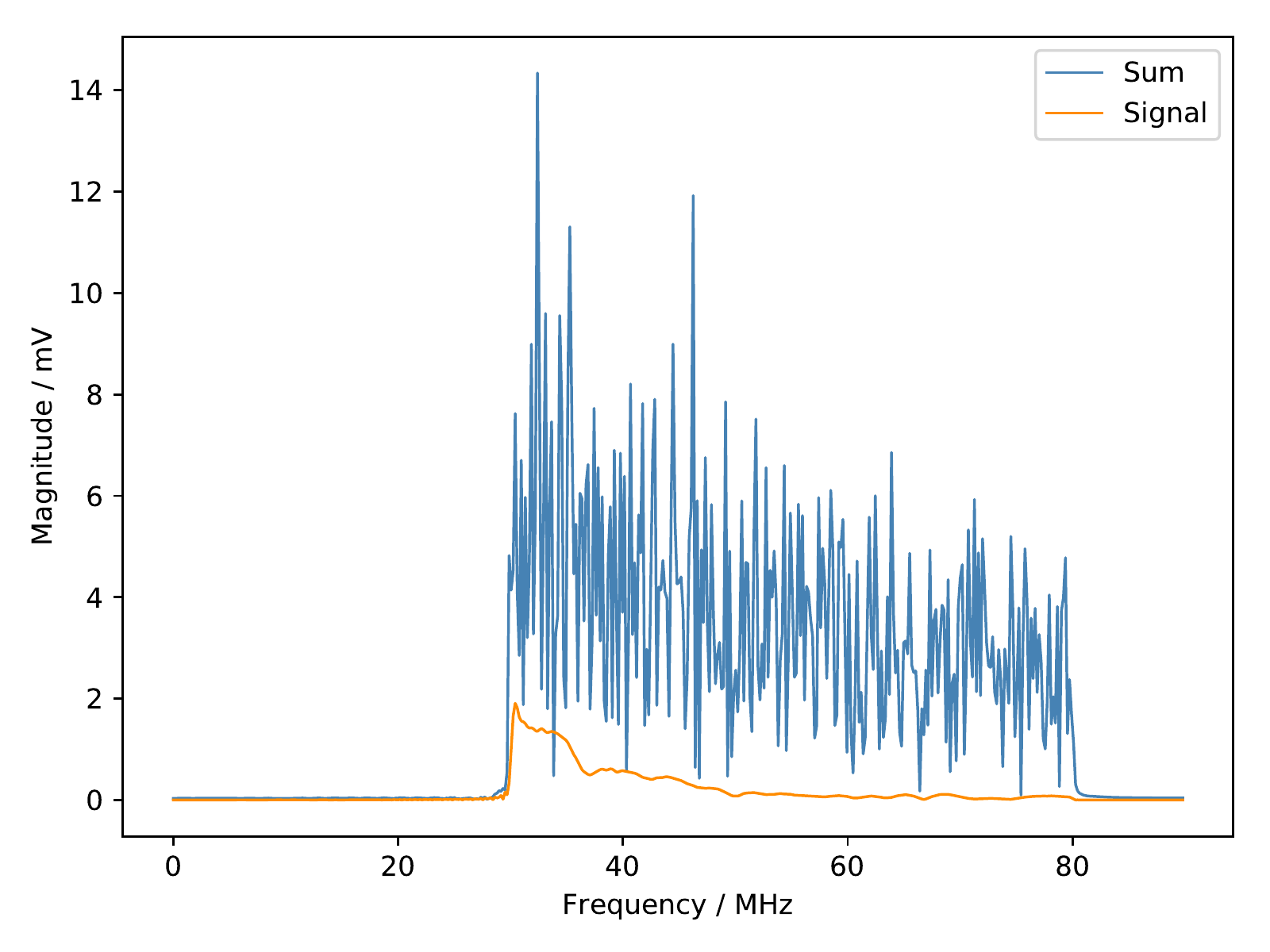}}
    \caption{Simulated trace with an air shower signal (orange) and the sum of the signal and noise (blue). The signal-to-noise ratio is $1.5$ in this case. The time trace and spectrum are shown in the top and bottom figure, respectively.}
    \label{fig:SimulatedTrace}
\end{figure}

We show an example of a simulated radio signal superimposed with simulated noise in Fig.~\ref{fig:SimulatedTrace}. The air shower radio signal in this example has SNR=$1.5$. Both the time trace and spectrum include the air shower signal and the sum of the signal and noise. The air shower signal exhibits a falling frequency spectrum, but the exact shape depends on the geometry of the shower.

\paragraph*{Data for signal recovery\\}

To present a realistic test scenario, we created a data set with a natural distribution of SNRs, with no scaling performed to achieve a specific SNR. The resulting SNR distribution peaks at 3.8 (median). On the other hand, SNR values for pure noise have a mean $\langle \text{SNR} \rangle = 2.7$ and standard deviation of $\sigma_{\textrm{SNR}}=0.5$, if we use the maximum noise amplitude in a random signal time window for $A_{\text{max}}^S$ in Eq.~\ref{eq:snr}.

To ensure that the network learns the shape of air shower signals, we consider only traces with the true maximum signal amplitude exceeding $30\%$ of the noise level (represented by the Root-Mean-Square, RMS) in a time window outside the signal region.

All traces are normalized to a maximum absolute amplitude of $1$ prior to their input to the network. Large differences in the signal strength are compensated with this normalization, and distinct features are retained for low signals. The network is trained on $69,967$ traces, while $7,775$ traces are used for validation. The network is then evaluated on an independent test set of $5,376$ traces with a SNR above $\sqrt{10}\approx 3.2$. We want to point out that no cut on the SNR was applied at either the training or validation stage.

The position of a true signal is identified by the maximum signal amplitude, and appears in a $2.2\mu\textrm{s}$ window within a trace. With a typical signal width of several hundred nanoseconds, this window mimics a signal search window for triggered data. This window is much longer than a typical electric field pulse of an air shower, which lasts only a few tens of ns, as the measured voltage is extended by dispersion effects in an antenna.

\section{Classification of signal and background events}
\label{Classification}

The primary challenge of an array of self-triggering antennas measuring cosmic rays resides in discriminating a signal from background in a continuous time trace. Our goal is to develop a deep neural network with efficient signal identification capabilities and strong background rejection.

\subsection{Basic considerations}
In our classification task, we analyze traces of $5.6\,\mu$s with $1,000$ bins, as is the case for AERA antennas. We require a trigger rate below $200$~Hz, while the false positive rate (FPR) must be below $0.1\%$. Therefore, we will search for neural network models having less than $20$ false triggers in the $20,000$ traces used for testing. Models will then be evaluated by their true positive rate (TPR), i.e. the percentage of correctly identified air shower events in the test data.

The SNR can be viewed as the primary obstacle to be overcome by a neural network. Therefore, our basic strategy is to train and test models for eight different values of the SNR, ranging from $0.5$ to $5$. In every training and testing cycle, the SNR value was kept constant.

\subsection{Network concept and training}

A neural network takes $1,000$ bins of a time trace as its input and delivers a binary output to the question "Is there a signal in this trace?". TensorFlow~\cite{tensorflow} was used in conjunction with the Keras interface~\cite{Keras} to set up and train our networks.

Initially, a search for suitable network architectures was performed, including extensive training and testing of network models together with a random search for optimal hyper-parameters. Inspired by the paper~\cite{arXiv:1701.00008}, where a convolutional neural network (CNN) was used to search for gravitational-wave signals in noisy data, we began exploring the same architecture. Nevertheless, two other sequential neural network architectures were compared to the CNN one. The first of these, long short-term memory (LSTM), required significantly more computational resources than CNN, and the results of the second architecture using only fully-connected (Dense) layers was not superior to CNN. Therefore, we decided not to pursue either of these architectures.

\subsubsection*{Architecture}

CNN has two components, convolutional (Conv) layers responsible for feature extraction, and Dense layers serving as a classifier. We tested various configurations and combinations of these layers, and our findings are described below. The number of hyper-parameters in tested models ranged from ten thousand to several million, and a random initial weight was supplied for each hyper-parameter. In addition, the number of filters in Conv layers, and the dimension of Dense layers, were drawn as $2^n$, where $n$ is a random number between one and eight. 

During our testing, we determined that results obtained from models with two Conv layers were better than those obtained from models using only one Conv layer. The next step was separating Conv layers into either two or three blocks with non-linearity and pooling layers following each block. On average, models with two blocks of layers outperformed models with only one block, and no improvement was found for models with three blocks. In addition, models with three Dense layers led to more correct results than those with one, two, or four layers.
If the performance of two models was similar, we chose the simpler of the two.

\begin{figure}[ht]
    \begin{center}
        \includegraphics[width=0.5\textwidth]{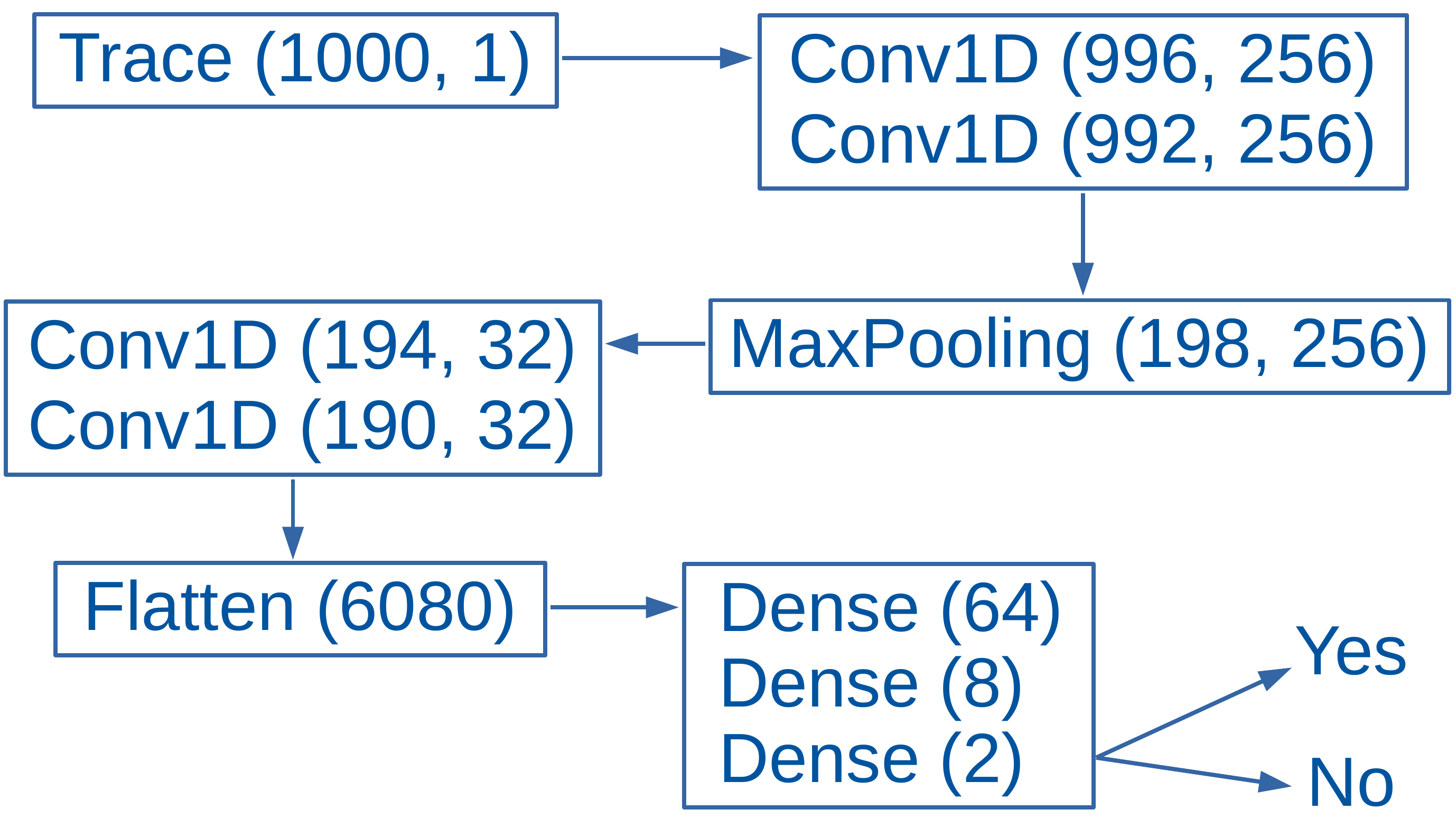}
        \caption{\it The classification model architecture. Dimensions are in parentheses. The input is a time trace with $1,000$~bins and the output is either yes or no for the presence of an air shower signal.}
    \label{fig:classification}
    \end{center}
\end{figure}

An overview of the best performing network model is shown in Fig.~\ref{fig:classification}. It consists of four one-dimensional convolution layers (Conv1D) separated into two blocks, and three fully-connected (Dense) layers. The initialization function draws samples from a truncated normal distribution centered on zero, called He\_normal~\cite{prelu}, and the Rectified Linear Unit (ReLU) activation function is used. The stride length of the convolution is unity. The ReLU activation function is used in all but the final layer. In the latter, the softmax function is adopted and works as a classifier. The model is trained with the Adam optimizer~\cite{1412.6980}, and the binary cross-entropy loss function.

The first two Conv1D layers have $256$ filter layers (i.e. the number of output filters in the convolution) and a kernel size of $5$, which specifies the length of the 1D convolution window. The second two Conv1D layers have their filter size reduced to $32$, with the number of filter layers being the same as that of the previous two layers. Batch normalization is performed after each convolution layer. The maximum pooling layer is used after the first block of convolution layers and the flattening layer after the second block in order to get the data into a format suitable for the following fully-connected layer (see Fig.~\ref{fig:classification}). This network has approximately $766,000$ trainable parameters.

\subsubsection*{Training}
 
We trained models using $100,000$ simulated traces for each SNR value. The training data set contained $50\%$ traces with a signal, with the other $50\%$ being traces with background only. An independent data set with $20,000$ traces was used to test the trained models. In this verification dataset, only $1\%$ of traces contained a signal, with this fraction being closer to realistic conditions in air shower experiments. The maximum of a radio signal was located in a random bin and was always present in a trace, but the same did not hold for other parts of the signal.

We found that learning did not improve after only a few epochs, as was found also for the signal recovery task (see Fig.~\ref{fig:training}), and this indicated sufficient training of models with the provided training datasets. We therefore decided to use only a single epoch in order to prevent over-training and reduce computation time. This approach was considered sufficient for the purpose of the classification task, as only Monte Carlo generated traces were used.

Training and testing were performed for SNR values between $0.5$ and $5$. About ten thousand training and test runs were done for each SNR value, allowing a sufficient scan of the hyper-parameters. The objective of our task was to increase the TPR while keeping the FPR below $0.1\%$. The best performing models were then saved for each SNR. We used the VISPA computing cluster (using NVIDIA 1080 GTX cards) at the RWTH Aachen University~\cite{Bretz:2012fu, vanAsseldonk:2015xng, Erdmann:2017xqb, Erdmann:2014vaa, VISPA-web} for our calculations.

\subsection{Classification results and working point}

\begin{figure}[ht]
    \begin{center}
        \includegraphics[width=0.7\textwidth]{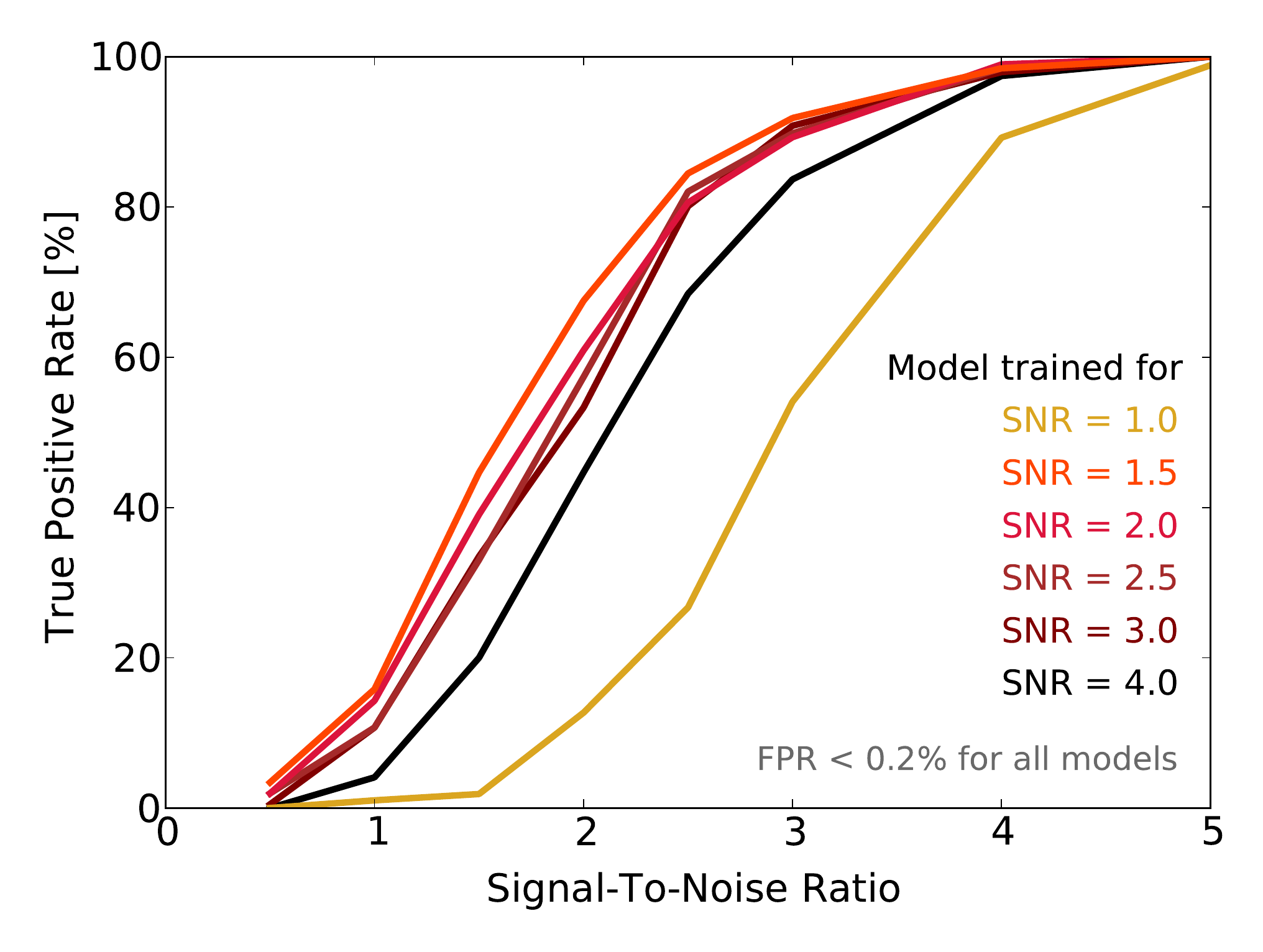}
        \caption{\it True positive rates obtained for models trained for the signal-to-noise ratio (SNR) indicated in the text, after being applied to other SNRs. The results for six selected models trained for SNRs ranging from $1.0$ to $4.0$ are shown with colored lines. All models had a false positive rate (FPR) below $0.2\%$ for any SNR.}
    \label{fig:classification-SNR}
    \end{center}
\end{figure}

We performed several checks of the best network models, and our results are summarized in Fig.~\ref{fig:classification-SNR}. The first step was a search for the best performing model for a given SNR. We made thousands of training runs for each SNR and saved the best models for all SNRs but SNR=$0.5$. The best performing model was defined as the model with the highest true positive rate (TPR) also having a false positive rate (FPR) below $0.1\%$.

In our second investigation we applied each of these six best models to time traces with SNRs different from the one used for the model training. Our goal was to get the TPR and also FPR at the other seven SNR values. The colored curves in Fig.~\ref{fig:classification-SNR} denote the obtained TPR for all six models as a function of SNR.

Generally, models trained for $1<$\,SNR\,$<4$ have good results, while the two remaining models have significantly lower TPRs. The best result for the classification was obtained with the model trained for SNR=$1.5$. The TPR values obtained with this model surpassed all other models for SNR\,$<$\,$4$. This model reached 45\%, 68\% and 84\% for SNR=$1.5$, $2$ and $2.5$, respectively. An important conclusion is that the FPR stayed below $0.2\%$ for all six models even for SNRs on which these models were not trained.

\section{Signal recovery}
\label{Signal-recovery}

Once a radio trace with an air shower signal has been recorded, the task is to extract the signal from a noisy trace. The recovery of radio signals is challenging, as they appear not to be very different from the ambient noise. We exploit the capabilities of deep learning techniques to differentiate between fine details in the properties and characteristics of signal and background traces.

\subsection{Strategy}

Our signal recovery aims at the reconstruction of air shower signals, i.e., the complete time traces with $1,000$~time bins (cf. Fig~\ref{fig:SimulatedTrace}, top). The air shower signal is fully contained in the time trace in this case. For this purpose we developed a network for regression, and used supervised training to reconstruct the entire signal from a noise-contaminated measurement. To train the network, simulated signal traces serve as labels, and the mean square error metric is utilized to calculate the loss function.

For a successful reconstruction it is crucial that the original signals are reconstructed as precisely as possible while suppressing noise. To asses that the traces were efficiently cleaned and the signals conserved, both the SNR (cf. Eq.~\ref{eq:snr}) and the signal energy contained in the traces
\begin{equation}
  \label{eq:trace_energy}
  E_{\text{signal}} = \frac{\Delta t}{R \cdot e} \, \left(\sum_{t_1}^{t_2} U_i^2 - \frac{t_2 - t_1}{t_4 - t_3} \sum_{t_3}^{t_4}  U_i^2 \right)
\end{equation}
are determined. The signal energy is calculated from the voltage trace $U(t)$ in units of electron volts (eV) using the elementary charge $e$, the antenna impedance $R=50\,\Omega$, and sampling rate $\Delta t=5.5$~ns. The noise contribution contained in the time interval $[t_3, t_4]$ is subtracted from the total energy in the signal window $[t_1, t_2]$ to recover only the signal energy.

\begin{figure}[t]
    \begin{center}
        \includegraphics[width=0.68\textwidth]{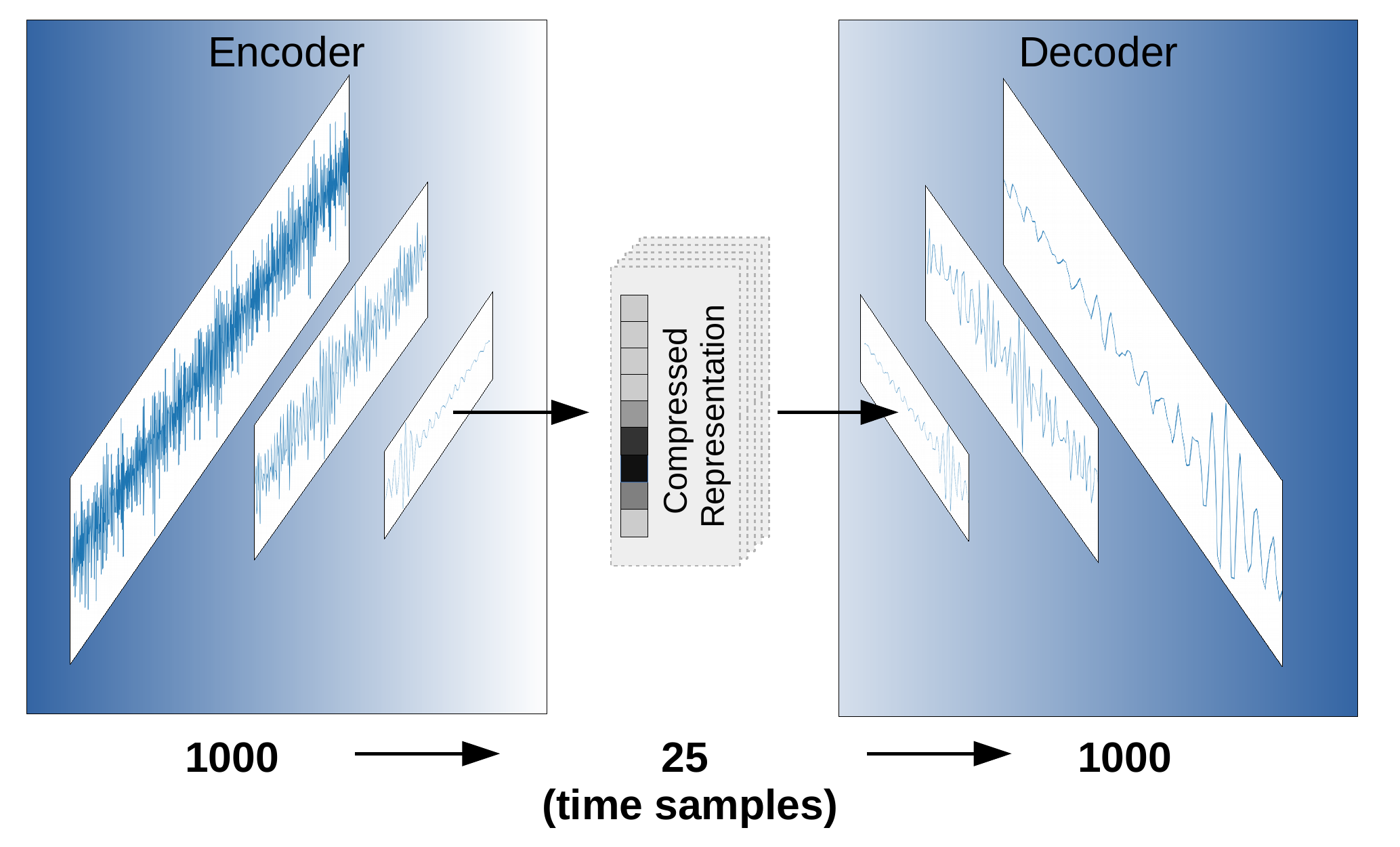}\hfill
		\includegraphics[width=0.28\textwidth]{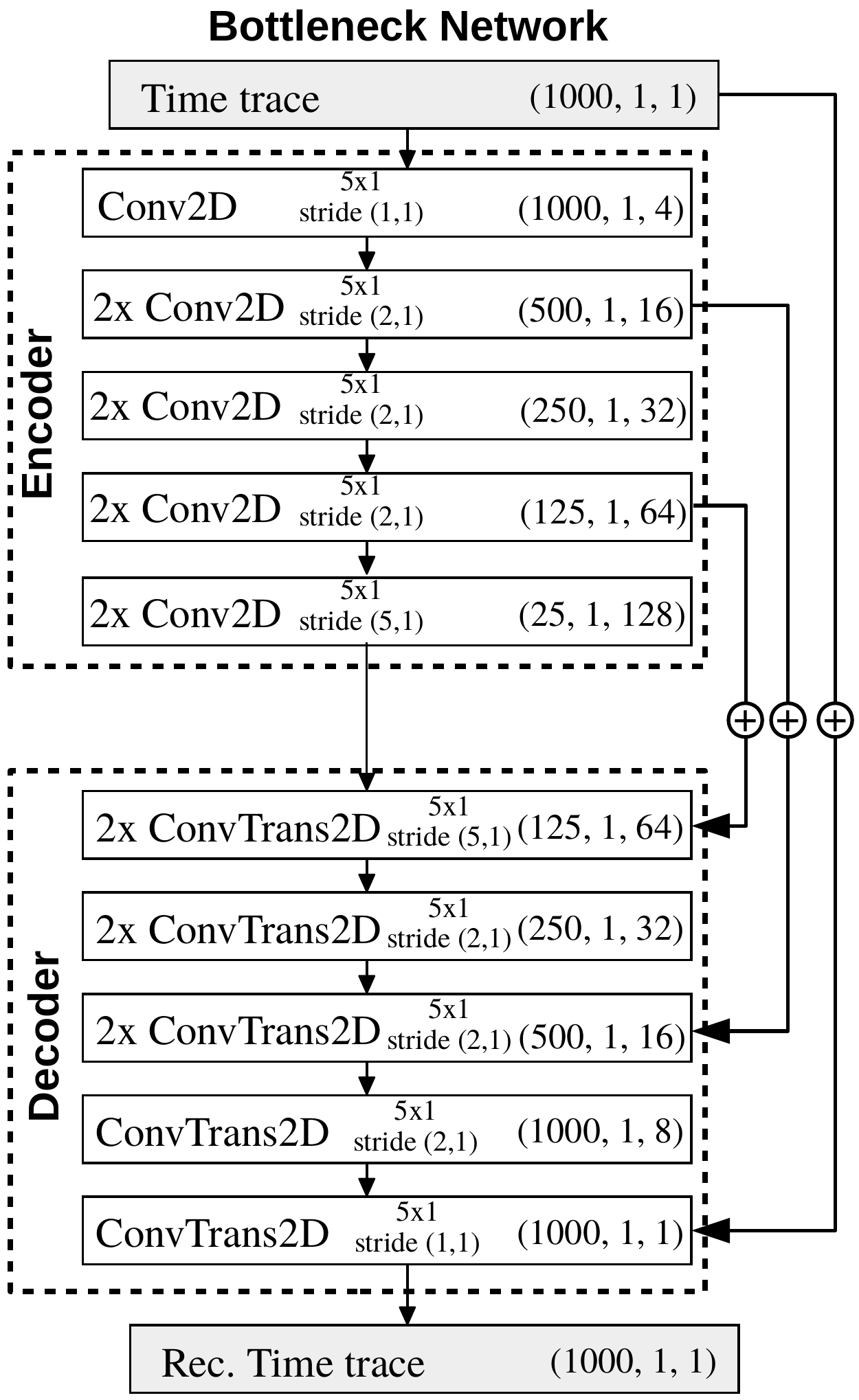}
        \caption{\it (Left) Illustration of a bottleneck-like network structure. First the input trace is compressed and multiple representations are created - encoding. Then these compressed representations are unfolded and one time trace is recovered - decoding. (Right) A detailed description of the de-noising autoencoder network used to recover air shower signals. The temporal dimension is reduced or increased with the stride value in the decoder or encoder part, respectively. All convolutional layers use a kernel size of 5$\times$1. The last three dimensions of the outgoing data are illustrated for each block on the right side. In the decoding part, transposed convolutions are performed.}
    \label{fig:autoencoder-principle}
    \end{center}
\end{figure}

\subsection{Network concept and training}
For implementation of the networks we once again used TensorFlow~\cite{tensorflow} with the Keras interface~\cite{Keras}.

The de-noising autoencoder network utilizes convolutional layers searching for translational invariant patterns in the one-dimensional time series. The architecture features a bottleneck-like structure illustrated in Fig.~\ref{fig:autoencoder-principle} (left). This structure consists of two parts: in the first stage, the trace is encoded (encoder) by decreasing the temporal dimension of the traces to multiple compressed representations. These compressed traces represent the input trace in terms of different patterns. In the second stage, the representations are decoded (decoder) and combined in order to recover the original dimensionality. In the whole procedure, the network searches for relevant features within the traces and subsequently chooses and unfolds only features associated with the signal.

This bottleneck design is utilized in other networks such as the so-called U-Net \cite{u_net} or unsupervised autoencoders. As an example, this concept has been studied for its applications in gravitational wave analysis~\cite{2017arXiv171109919S}. An advantage of the bottleneck feature is that with comparable performance, the number of free parameters (weights, biases) is reduced relative to other network designs.

\subsubsection*{Architecture}

Within the network, the temporal dimension is initially decreased from $1,000$ to $25$ bins and afterwards restored. A detailed description of the network's layout is presented in Fig.~\ref{fig:autoencoder-principle} (right). To decrease the temporal dimensionality, striding of $2$ or $5$ time bins is used within the convolutional layer. All layers have a fixed kernel size of $5$ bins. 

First, a single layer with four filters takes the input data without changing its dimensionality. The reduction is then performed in four blocks, each consisting of two layers according to the reduction sequence of $1000 \rightarrow 500 \rightarrow 250 \rightarrow 125 \rightarrow 25$ time bins. In each block, the first layer decreases the dimension while the second preserves the size. Both layers use the same number of filters, which doubles for each block from $16$ to $128$. In the same way, the decoder unfolds the trace again. The encoding layers utilize 2D convolutions, while the decoding layers perform transposed convolutions~\cite{Keras}.

Additionally, three shortcuts are realized between several layers which superimpose the features represented by the layers (cf. Fig.~\ref{fig:autoencoder-principle} right). With these connections, gradient back-propagation is supported, resulting in better training stability as discussed in so-called Residual Networks~\cite{res_net}. In our network, the shortcuts were essential for successful training.

As we aim to reconstruct oscillating time series, the activation function needs to cope with positive and negative values. The parametric rectified linear unit PReLU~\cite{prelu} enables negative values, and has been used instead of the widely used ReLU activation. We used $128$ as the batch size.

Tests regarding the architecture with increased complexity, i.e., increased number of filters or layers, revealed no improvement in performance. As an alternative architecture, we also investigated a convolutional network keeping the temporal dimension of the traces within the network. No major difference in the performance of the signal recovery was observed. Owing to fewer parameters, the presented bottleneck network required ca. $30$ sec per epoch on the NVIDIA 1080 GTX card, which turned out to be $30-40\%$ faster than the convolutional network. Thus we report the results of the more computationally efficient bottleneck network in this paper.

\subsubsection*{Training}

Here we also use the Adam optimizer~\cite{1412.6980} for our supervised network training, with a learning rate of 0.001. We present to the network simulated traces of air shower signals superimposed on quasi-realistic noise traces as the input and give the pure signal traces as labels, which should be recovered by the network from the input traces.

\begin{figure}[t]
    \begin{center}
        \includegraphics[width=0.5\textwidth]{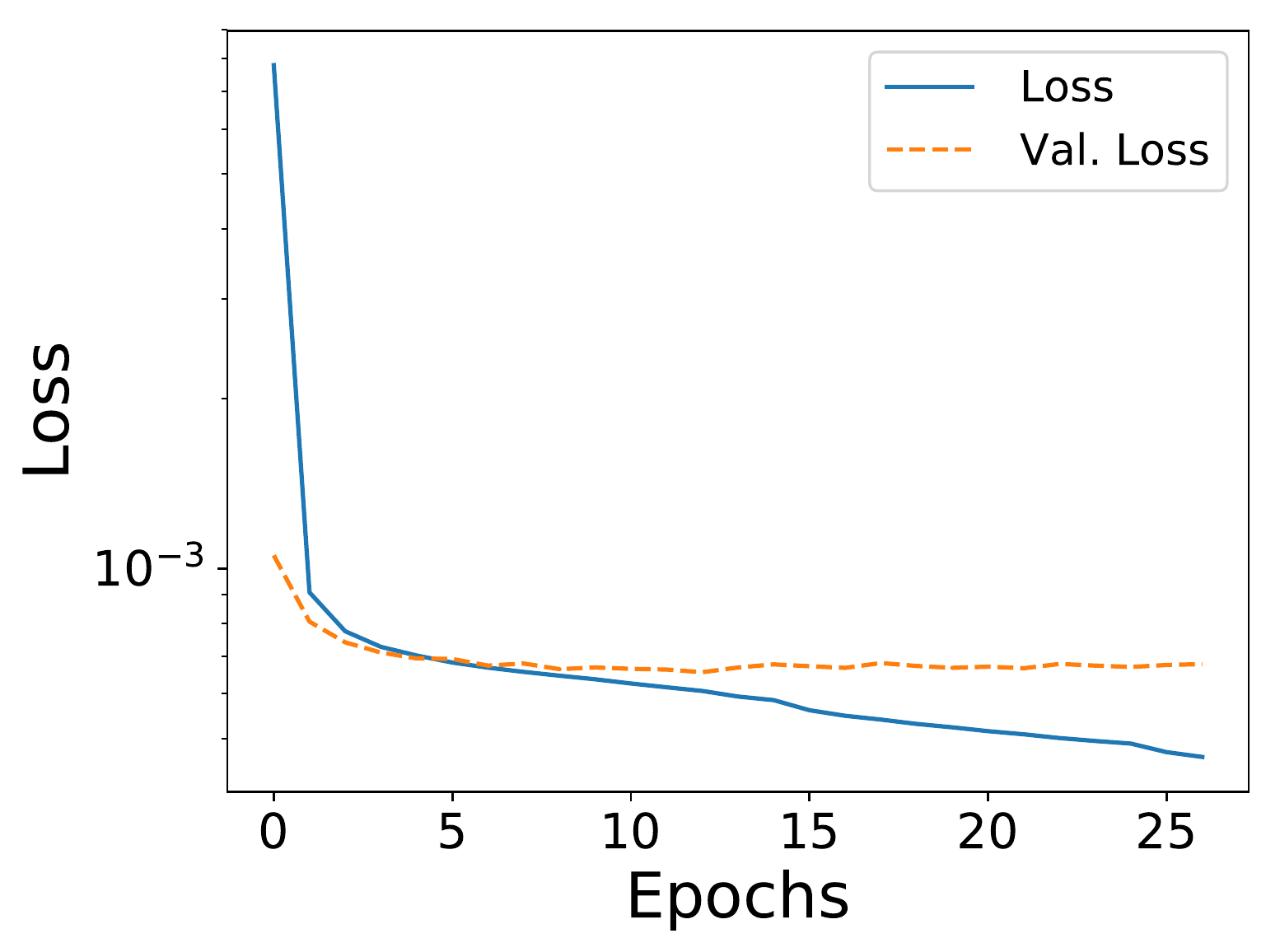}
        \caption{\it Training and Validation loss as a function of the epoch.}
    \label{fig:training}
    \end{center}
\end{figure}

The quality of the training is described by the loss functions (here the mean square error metric) for the training and validation data sets after each epoch. The corresponding curves, presented in Fig.~\ref{fig:training}, initially show a fast decrease, indicating effective learning. The training loss then further decreases (note the logarithmic scale), indicating sufficient complexity of the network for learning, while the validation loss stagnates without signatures of over-training. The training is stopped after 26 epochs using the Keras callback "EarlyStopping" \cite{Keras}, as no further improvement in terms of validation loss is achieved.

\subsection{Energy of the air shower signal}

\begin{figure}[t]
    \begin{center}
        \includegraphics[width=0.95\textwidth]{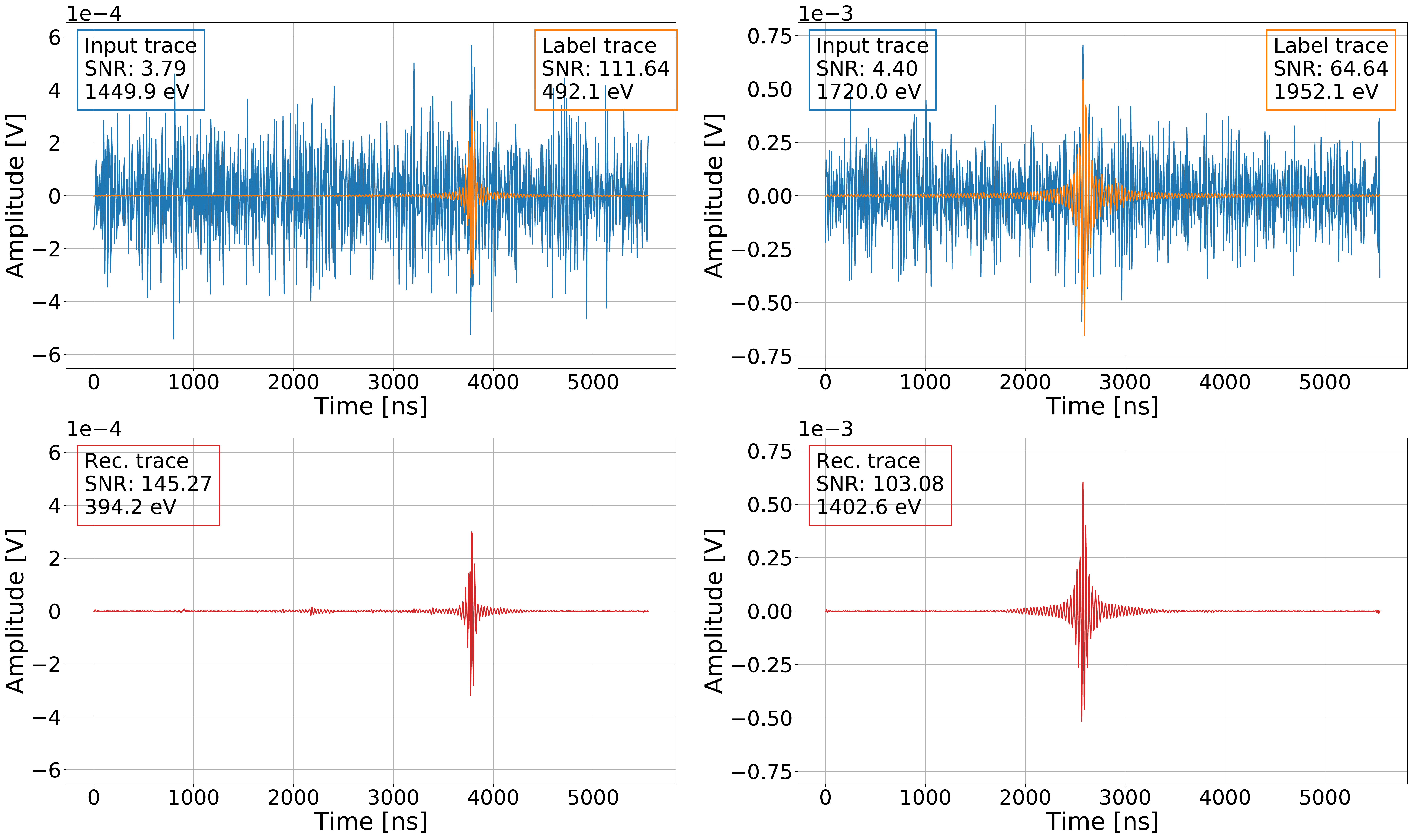}
        \caption{\it Two examples of the signal recovery and reconstruction by the network. Input traces with simulated signal superimposed with noise (blue) and only true signal trace (yellow) are at the top. Recovered signal trace (red) is at the bottom. The SNR and signal energy for each trace is shown in the colored boxes. Signals appear in a 2200~ns window centered at 3000~ns.}
    \label{fig:reco}
    \end{center}
\end{figure}

The trained network can be used to recover signals from traces contaminated with noise. Two examples of signal recovery with the de-noising autoencoder are presented in Fig.~\ref{fig:reco}. The upper two figures show the pure radio signal (yellow traces) and its superposition with noise (blue traces).

The lower two figures show the corresponding reconstructed signal traces (red) on the output of the network. In both examples, the network correctly identifies the signal and reconstructs a signal trace with no significant noise contribution. As a result of the signal reconstruction, the SNR increases significantly for the reconstructed traces (cf. the upper left boxes in the figures show the SNR values and the signal energy in units of eV). The reconstructed signals have the proper shape and amplitude. The deviation in the signal energies improves compared to the input traces. For example, the signal energy decreases from 1.45~keV to 0.39~keV while the true (label) signal has 0.49~keV for the event shown on the left in Fig.~\ref{fig:reco}.

The network performance is also evaluated on the entire test data set. First, the signal recovery is examined. To assess whether the signal contributions were properly recovered, the deviation in energy is calculated as
\begin{equation}
\frac{\Delta E_\textrm{i}}{E_{\textrm{true}}} = \frac{E_{\textrm{i}} - E_{\textrm{true}}}{E_{\textrm{true}}},
\end{equation}
where $E_{\textrm{true}}$ is the energy obtained from the label traces according to Eq. \ref{eq:trace_energy}. $E_\textrm{i}$ donates the signal energy obtained from a trace recovered with the network (i$\,$=$\,$rec) or taken directly from the input traces (i$\,$=$\,$input).

\begin{figure}[t]
    \begin{center}
        \includegraphics[width=0.5\textwidth]{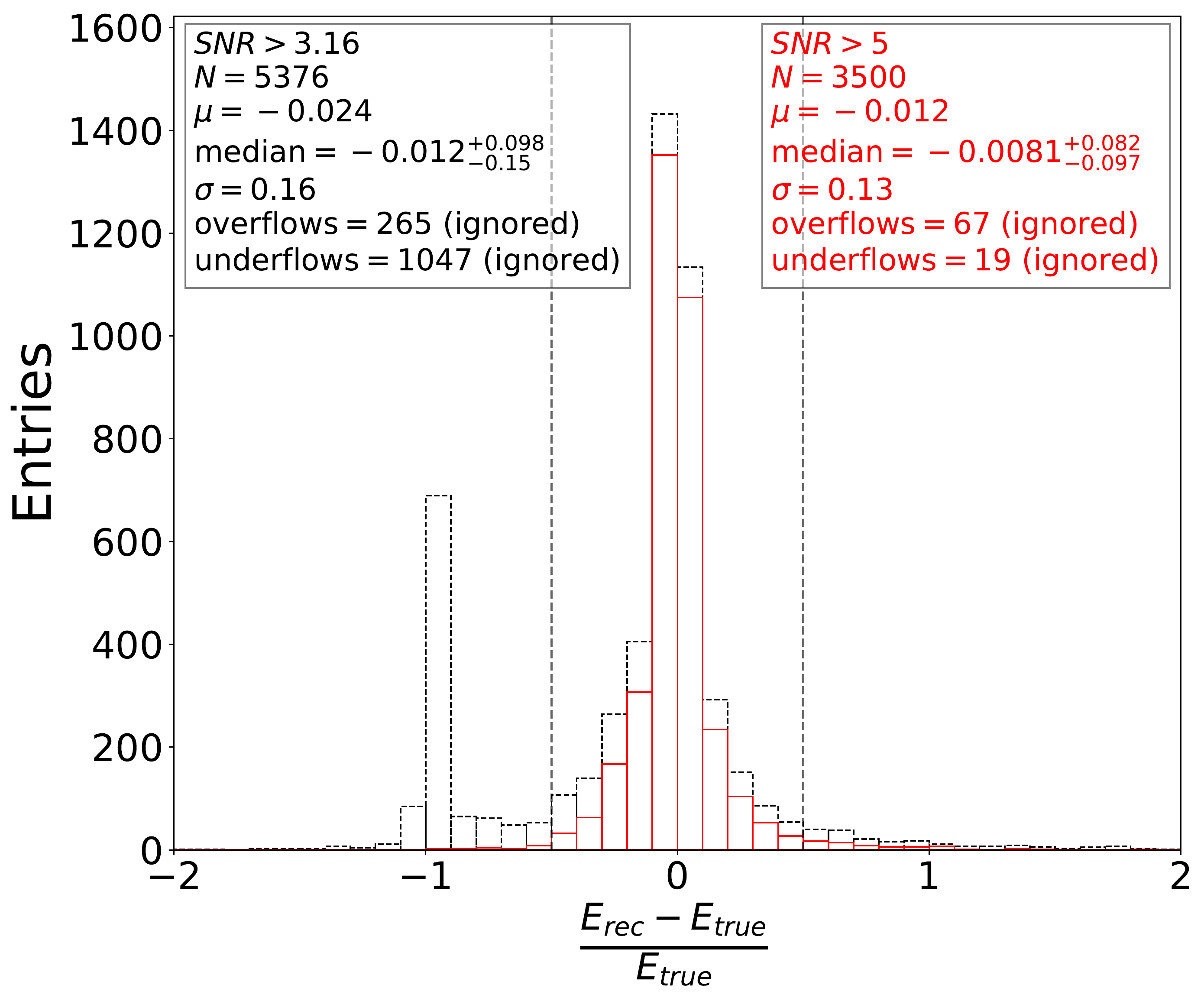}
        \caption{Histogram of the normalized energy deviation between recovered and true signals. Energy is obtained according to Eq.~\ref{eq:trace_energy}. Values beyond $|\Delta E|/E_{\textrm{\normalfont true}}>0.5$ are excluded for the calculation of the mean and width and are accounted in over- and under-flow. The dashed histogram contains all events in the test set with $\mbox{SNR}_{\textrm{\normalfont input}}>3.16$, while the red histogram resembles the distribution for all events with $\mbox{SNR}_{\textrm{\normalfont input}}>5$.}
    \label{fig:energy}
    \end{center}
\end{figure}

The distribution of the energy deviation of the recovered traces is presented as a dashed histogram in Fig.~\ref{fig:energy}. It has a prominent peak at the median, close to zero, and a width $\sigma\simeq20\%$.
The additional peak around $-1$ indicates that $\sim 20\%$ of the signals were not recovered. For these traces, the network reconstructed either a noise pulse or no significant signal at all. All of these traces have low SNR and hence, low signal. This is verified by the red histogram in Fig.~\ref{fig:energy}, which shows the deviation in energy only for events with $\mbox{SNR}>5$.

\begin{figure}[t]
    \begin{center}
        \includegraphics[width=0.48\textwidth]{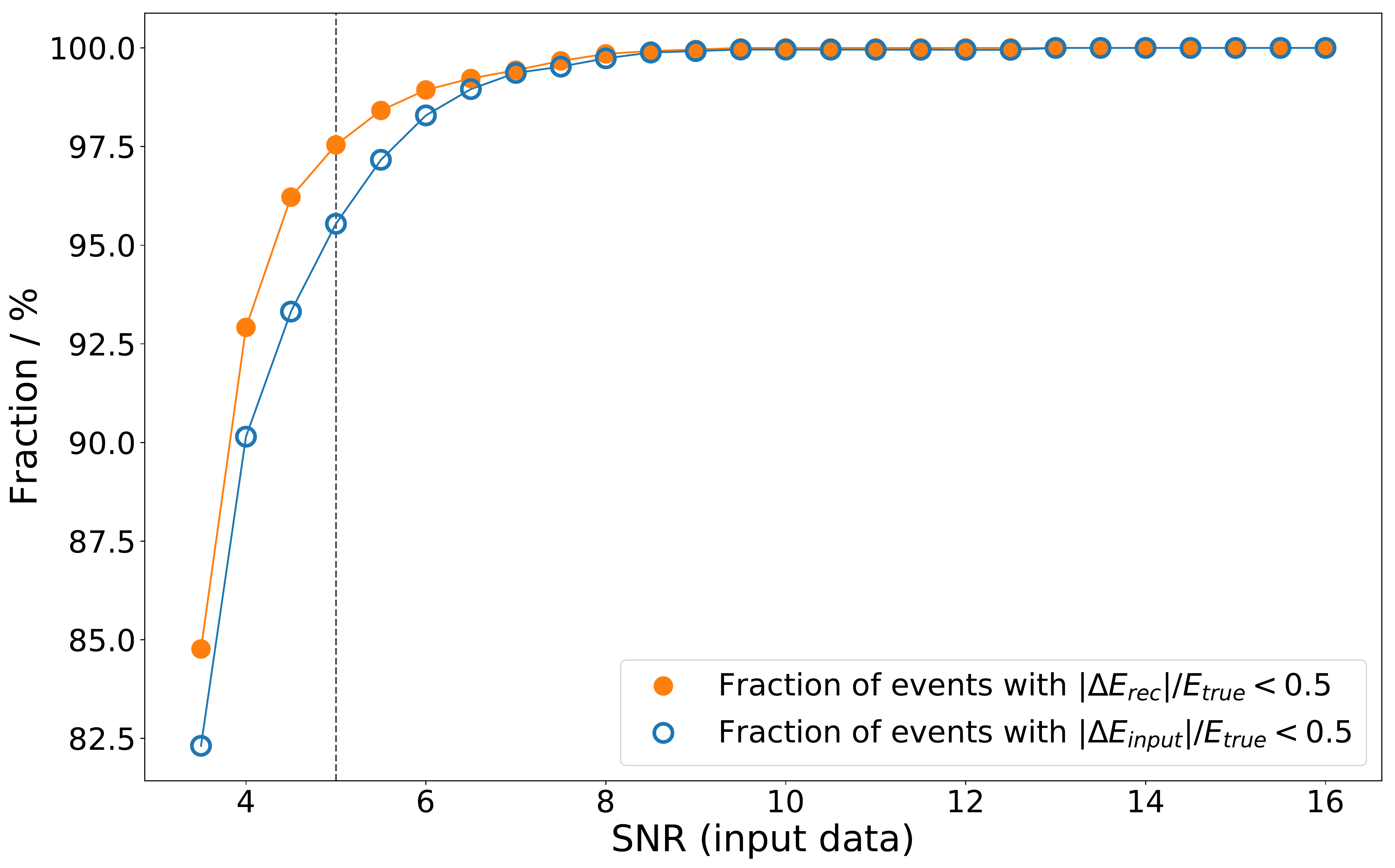}\hfill
        \includegraphics[width=0.48\textwidth]{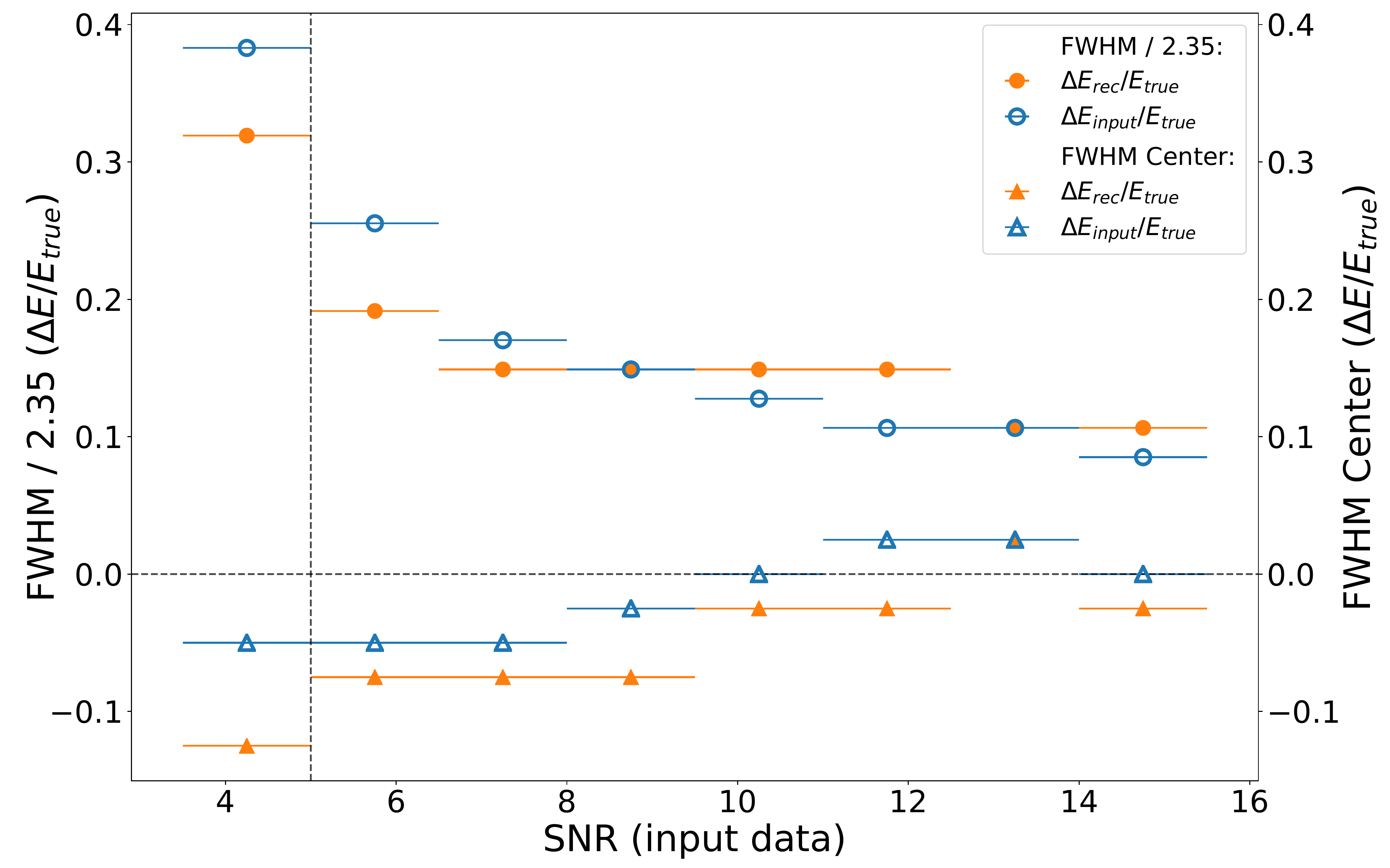}
        \caption{\it (Left) Fraction of events with $|\Delta E|/E_{\textrm{\normalfont true}}<0.5$ as a function of the minimum $\textrm{SNR}_{\textrm{\normalfont input}}$.
        (Right) The energy resolution and bias for the reconstructed signals in means of FWHM/2.35 (circle) and the FWHM center (triangles) determined from the distribution $|\Delta E|/E_{\textrm{\normalfont true}}$ in a given SNR interval (shown by the horizontal error bar). In both figures, the yellow and blue curve shows the result for signals recovered with the autoencoder network ($\Delta E_{\textrm{rec}}$) and determined from the input traces using Eq. \ref{eq:trace_energy} ($\Delta E_{\textrm{input}}$), respectively. The vertical dashed line marks $\mbox{SNR}=5$.}
    \label{fig:snr_energy}
    \end{center}
\end{figure}

The latter distribution indicates that the ability of the network to properly recover the signal depends on the SNR. The fraction of events recovered by the network with $|\Delta E\textrm{\normalfont rec}|/E_{\textrm{\normalfont true}} < 0.5$ as a function of the minimum SNR$_{\textrm{\normalfont input}}$ is shown as a yellow curve in Fig.~\ref{fig:snr_energy} (left). We can see that 97$\,\%$ and almost all events are reconstructed with $\mbox{SNR}=5$ and $8$, respectively. The blue curve shows the fraction of events with $|\Delta E_\textrm{\normalfont input}|/E_{\textrm{\normalfont true}}<0.5$ when calculating the signal energy from input traces using Eq.~\ref{eq:trace_energy}. This equation compensates for a certain noise contribution by subtracting the energy deposit in a signal window with the content from a noise window. The same strategy is used for the data measured by the AERA array \cite{AERA-Energy-PRD}. Comparing the two curves in Fig.~\ref{fig:snr_energy} reveals that more low SNR events are reconstructed using the autoencoder network.

We determine the Full Width at Half Maximum (FWHM) from the histogram of reconstructed signals $\Delta E / E_{\textrm{true}}$ within a given interval of SNR. The FWHM$/2.35$ serves as a measure of the width and the FWHM center as the mean of the distributions and are shown as a function of the SNR in Fig.~\ref{fig:snr_energy} (right). In order to have a robust and reliable estimation of the energy resolution and bias at low SNR, we choose this metric over a normal distribution. We get comparable results with the FWHM and normal distribution for high SNR.

As before, the yellow markers identify the resolution (circles) and bias (triangles) of signals recovered by the network while the blue markers are determined by using Eq.~\ref{eq:trace_energy} on the input traces. 

The network improves the resolution of the signal energy for low SNR. The resolution for events with the SNR between 5 and 6.5 increases from 26\% to 19\%. This improvement slows down until the resolution determined directly from the input traces is better at $\text{SNR}\sim10$, where the resolution is below 15\%. Let us note that the designed deep learning algorithm can be adjusted, e.g. by the choice of the normalization, to increase resolution for low-SNR events while only slightly worsening results for high-SNR events. This study is dedicated to improving the reconstruction for low-SNR events as for high-SNR events the trace cleaning becomes unnecessary.

The FWHM center reveals a bias in the signal energy reconstruction for both approaches, see Fig.~\ref{fig:snr_energy} (right). The reconstruction of the energy for signals recovered by the network shows a bias of $\sim10\%$ for low SNR and this bias vanishes for $\text{SNR}>10$. The bias at low SNR will contribute to the total systematic uncertainty, which will be still smaller than the total systematic uncertainty of 28\% reported for the square of the reconstructed electric-field amplitudes for AERA \cite{AERA-Energy-PRD}.

We can conclude that the autoencoder network improves the recovery of cosmic ray radio pulses in a single-polarization measurement, particularly in the case of weak signals. A direct comparison of the network with standard techniques applied to real measured data, including electric-field traces of at least two polarization channels, would be beneficial and could lead to improvements of the network.

\subsection{Frequency spectrum of the air shower signal}

\begin{figure}[t]
    \begin{center}
        \includegraphics[width=0.6\textwidth]{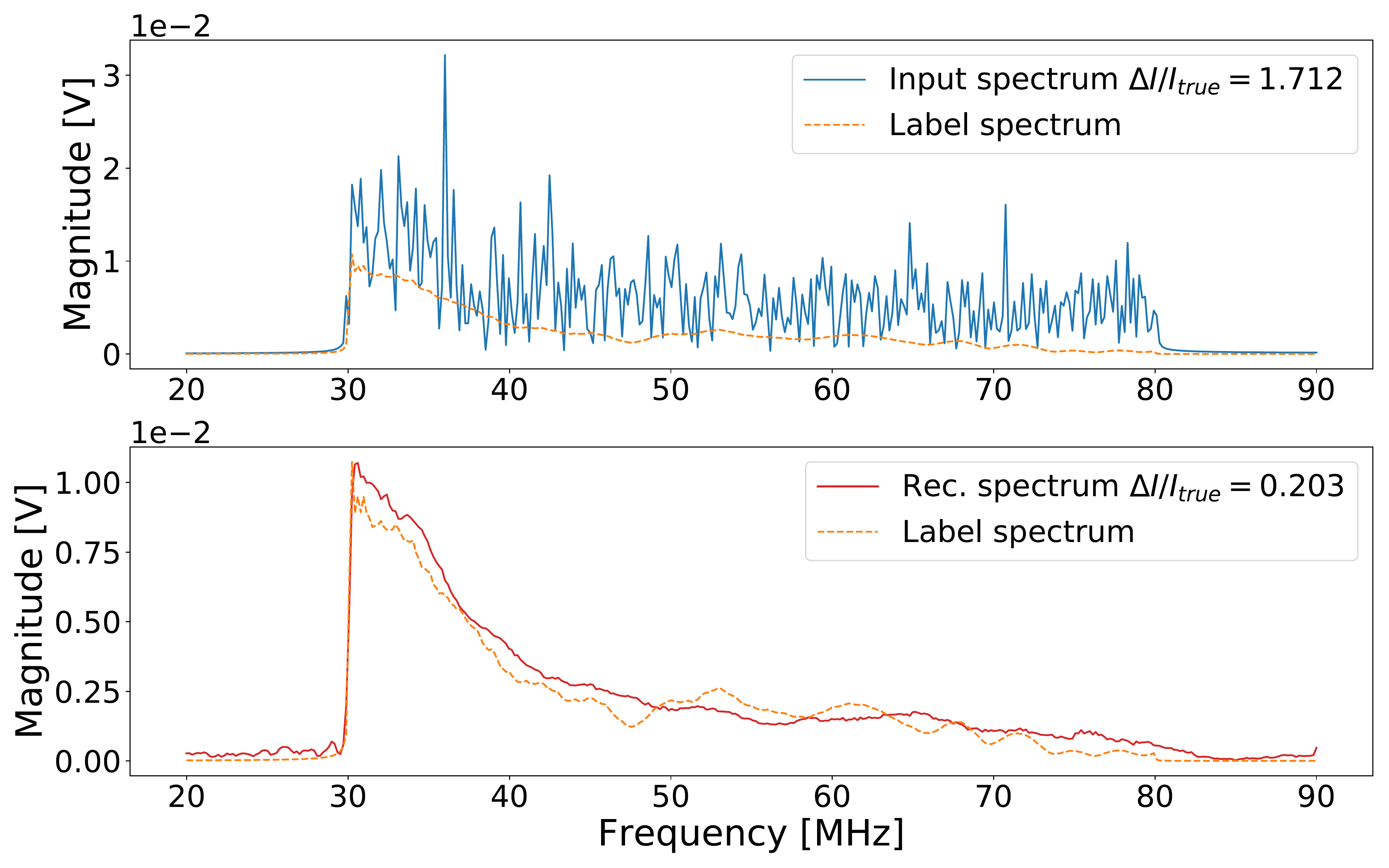}\hfill
        \includegraphics[width=0.36\textwidth]{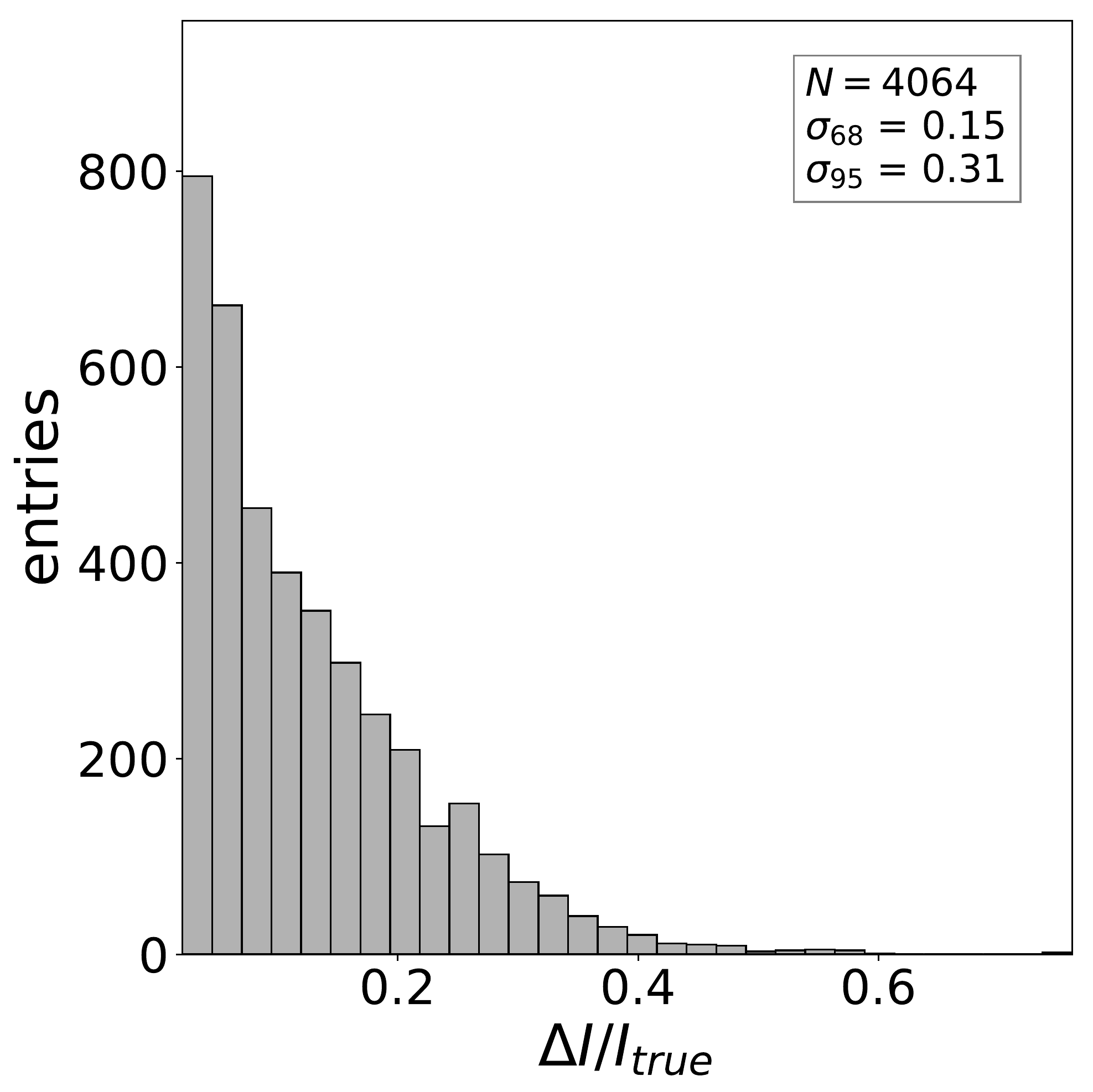}
        \caption{\it (Left) Frequency spectra of an input trace (blue, top), a label trace (yellow, top and bottom), and reconstructed trace (red, bottom).
        (Right) Distribution of $\Delta I/I_{\text{true}}$ after Eq.~\ref{eq:frequency} for all traces with $|{E_{\text{rec}}-E_{\text{true}}}|/{E_{\text{true}}}<0.5$ (cf. Fig.~\ref{fig:snr_energy} (right)).}
    \label{fig:frequency}
    \end{center}
\end{figure}

We also compared the frequency distribution of the reconstructed and label time traces. The frequency spectrum of an event is shown in Fig.~\ref{fig:frequency} (left). The total signal and air shower signal, or the label, are shown as blue and yellow curves, respectively. The air shower spectrum exhibits a falling distribution in the filtered interval between $30$ and $80$~MHz. Noise contributions contained in the blue curve show large fluctuations.

The frequency spectrum of the signal reconstructed by the network (red) is shown in the lower left part of Fig.~\ref{fig:frequency} together with the label spectrum (yellow). In order to quantify the accuracy of the reconstruction in the frequency domain, we calculate the integral over the difference between reconstructed (input) and true signal spectrum normalized by the true signal spectrum
\begin{equation}
    \label{eq:frequency}
    \frac{\Delta I}{I_{\text{true}}} = \frac{\int_{30\,\text{MHz}}^{80\,\text{MHz}}|\mathcal{F} - \mathcal{F}_{\text{true}}| }{\int_{30\,\text{MHz}}^{80\,\text{MHz}}\mathcal{F}_{\text{true}}},
\end{equation}
where $\mathcal{F}$ is the absolute value of the complex spectrum. We have $\Delta I / I_{\textrm{true}} = 0.2$ for the presented example. The distribution of $\Delta I/I_{\textrm{true}}$ for all traces reconstructed with $|{E_\textrm{rec}-E_{\textrm{true}}}|/{E_{\textrm{true}}}<0.5$ (see Fig.~\ref{fig:snr_energy} (right)) is shown in Fig.~\ref{fig:frequency} (right). The distribution falls with a $68\%$ quantile of $\sigma_\mathrm{{68}}=0.15$ and a $95\%$ quantile of $\sigma_\mathrm{{95}}=0.31$. This investigation in the frequency regime confirms that the signal recovery from noise contaminated time traces is feasible. Note that the frequency spectra of the signals were not used
for the training of our network.

\section{Conclusion \label{sec:conclusion}}

In this work, we investigated deep learning methods for classification and reconstruction of radio signals emitted by cosmic ray induced air showers.

The identification of radio signals within a noise contaminated environment had a true positive rate of about $90\,\%$ for signal-to-noise ratios SNR$\,>\,3$, while the false positive rate was below $0.2\,\%$. This level of SNR is typically used as the minimum value for the reconstruction of measured radio signals. When including contributions from radio frequency interference (RFI) this picture does not change. The best model found was trained for signal amplitudes close to the noise level, i.e. SNR=$1.5$, that may be targeted by future experiments.

A rather accurate reconstruction of the identified radio signal can be achieved by a de-noising autoencoder; a deep neural network model adopting a bottleneck-like structure. The algorithm combines an encoding and unfolding of time traces, providing efficient noise suppression.
About $80\,\%$ of signals were recovered including heavily noise contaminated traces with low signal-to-noise ratios. This number increases to above 97$\,$\% for events with a signal-to-noise ratio beyond 5. From the de-noised traces the signal energy is calculated with a resolution of $\sim\,20\,\%$ for events with a signal-to-noise ratio between 5 and 6.5. This resolution improves with increasing signal-to-noise ratio. To benchmark our network we compared the fraction and resolution determined from de-noised traces with results obtained for the noise contaminated traces. This comparison reveals better results using the network for events with low signal-to-noise ratios.

A benchmark comparison in frequency-space revealed that RFI signals were effectively suppressed and the spectral shape of the signal was recovered well.

We presented two neural network models which successfully identify air shower signals and reconstruct the signal energy. These networks can be used in any current or future observatory~\cite{Auger, GRAND}. The authors foresee additional improvements in both analyses by using the signal measured by two channels in an antenna, or by exploring the information in the frequency spectrum of measured traces.

To further optimize the models, data collected at the site of an experiment could be used as an alternative to Monte Carlo-generated data. Training can improve classification and signal reconstruction by using measured noise signals. The simulation studies presented in this work show the potential of applying deep learning methods to the radio detection of air showers.

\acknowledgments

It is our pleasure to acknowledge the interaction and collaboration with many colleagues from the RWTH Aachen University and the Pierre Auger Collaboration. We are grateful to C. Glaser for sharing his simulated data sets and to J. Glombitza for his valuable comments on deep learning techniques. We thank an anonymous reviewer for her/his useful comments and M. Malacari for reading the manuscript. We acknowledge the financial support of the Ministry of Innovation, Science and Research of the State of North Rhine-Westphalia, and the Federal Ministry of Education and Research (BMBF).


\end{document}